\documentclass[12pt]{article}

\usepackage{setspace}
\usepackage{amsfonts}
\usepackage{amsmath}
\usepackage{amssymb}
\usepackage{color}
\usepackage{enumerate}
\usepackage{verbatim}
\usepackage{url}
\usepackage[dvips]{epsfig}
\usepackage{graphicx}
\usepackage{subfigure}

\newcommand{\sT}{S_{T}}
\newcommand{\sA}{S_{A}}
\newcommand{\dN}{D_{N}}

\title{A Measure of Control for Secondary Cytokine-Induced Injury of Articular Cartilage: A Computational Study}
\author{Jason M.\ Graham$^{1}$}

\begin{document}
\maketitle

${ }^{1}$ Department of Mathematics, University of Scranton, Scranton, PA, USA

\begin{abstract}
  In previous works, the author and collaborators establish a mathematical model for injury response in articular cartilage. In this
  paper we use mathematical software and computational techniques, applied to an existing model to explore in more detail how the behavior of cartilage cells is influenced by several of, what are believed to be, the most significant mechanisms underlying cartilage injury response at the cellular level. We introduce a control parameter, the radius of attenuation, and present some new simulations that shed light on how inflammation associated with cartilage injuries impacts the metabolic activity of cartilage cells. The details presented in the work can help to elucidate targets for more effective therapies in the preventative treatment of post-traumatic osteoarthritis.
\end{abstract}


\section{Background}

Injury response and wound healing is a topic of central importance in biomedical research for obvious reasons. As a result, there has been a great deal of activity in developing mathematical and computational models of wound healing in various organ systems such as skin, see e.g.\ \cite{sherratt1990}. In contrast, there has been very little activity in developing computational models for wound healing and injury response in articular cartilage, despite a great interest in this topic in orthopaedics research. What is more, few if any of the mathematical models developed for wound healing in other systems are appropriate for application to articular cartilage.

Articular cartilage is made up of differentiated mesenchymal cells known as chondrocytes. These cells are embedded in an extracellular matrix and are responsible for the biomechanical properties of cartilage \cite{ulrich2003}.  Mechanical stress and injury influence changes in the metabolic activity of chondrocytes \cite{ulrich2003}. Specifically, during injury response chondrocytes produce and respond to certain cytokines, or signaling molecules, known as tumor necrosis factor $\alpha$ (TNF-$\alpha$) and erythropoietin (EPO). There is a ``balancing act'' between the pro-inflammatory cytokine TNF-$\alpha$ and the anti-inflammatory cytokine EPO in which each limits the production and biological action of the other.
In a recent article Brines and Cerami \cite{brines2008} suggest that TNF-$\alpha$ plays a significant role in causing the spread of cartilage lesions, while EPO plays an antagonistic role to TNF-$\alpha$, limiting the area over which a lesion can spread by counteracting some of the effects of inflammation \cite{brines2008}. It has also been observed that there are inherent time-delays in the activation of, and signaling by EPO that results in a window of opportunity for the spread of lesions due to secondary injury caused by inflammation. However, the authors of \cite{brines2008} suggest that it may be possible to intervene with EPO derived therapies to minimize the amount of secondary injury due to inflammation and the spread of
cartilage damage.

In previous works \cite{graham1,graham2}, the author, with collaborators, develop a novel mathematical model for articular cartilage injury response aiming to test hypotheses put forth in \cite{brines2008}.  As with any mathematical or computational model, it is important to understand how the behavior of the results depend on the parameter values. In particular, it is useful to know how changes in the parameter values effect the results of simulations. This is especially the case when the goal is to tie the modeling efforts to experimental results. On the one hand, any measurement involves an error and if the model is sensitive to small changes in the parameter values, often the case when models contain nonlinear terms, the experimental error may be significant enough that simulations behave differently than what is to be expected, based on experimental observations. On the other hand, different parameter values may correspond to different types of observed behavior, or even more interestingly, changes in parameter values can lead to predictions about the system being modeled. One of the goals of the models described in \cite{graham1,graham2} is to understand the balance between pro-inflammatory cytokines such as TNF-$\alpha$ and anti-inflammatory cytokines such as EPO. We can use the mathematical models together with computational techniques to help understand this balance by exploring how changes in parameters related to different aspects of TNF-$\alpha$ and EPO dynamics simulate different types of behavior in cartilage injury response. This paper is devoted to such an exploration. In particular, we would like to know how changes in parameter values corresponding to different properties of the TNF-$\alpha$/EPO interactions influence the lesion expansion or abatement properties during cartilage injury response.

The remainder of this paper is organized as follows. The next section provides a brief description of a mathematical model, described fully in \cite{graham1,graham2}\footnote{We note that there is a slight difference between the models in \cite{graham1}, and in \cite{graham2}. In this work we use the model in \cite{graham1} as it gives the same (qualitative) results but replaces a discontinuous term with a continuous term, and also replaces a phenomenological parameter with one that is more directly connected to the biology.}, to which, in this paper, we apply computational methods to explore some issues regarding the behavior of chondrocytes during the typical injury response in articular cartilage. It is in that section where we establish ideas and notation that is used throughout the remainder of this work. The third section, the results section, shows the computational results and discusses their significance. The paper ends with conclusions
drawn from the results section.

\section{Materials and Methods}

Here we briefly summarize the mathematical model, established in \cite{graham1,graham2}, used to obtain the computational results of the next section.
During injury, chondrocytes are considered as being in specific and distinct ``states'' corresponding to which cytokines the cells are capable of producing and responding to. We refer to the normal state of a subpopulation of chondrocytes as the healthy state. As a result of inflammation and injury, healthy chondrocytes can enter into a ``sick'' class in which they are at risk of undergoing programmed cell death. The sick cells are considered as being in one of two states:
 \begin{enumerate}
   \item the catabolic state
   \item the EPOR active state
 \end{enumerate}
 Cells in the catabolic state are characterized by their ability to produce TNF-$\alpha$, while EPOR active cells are characterized by their
 ability to express a receptor for EPO. We note that these two cell states are distinct in that cells capable of producing TNF-$\alpha$ are
 not capable of expressing the EPO receptor, and vice versa. Another consequence of cells being in the catabolic state is that they produce
 reactive oxygen species (ROS) which serves as a catalyst for the production of EPO by cells in the healthy state.

 Due to the fact that there are two typical means of cell death: necrosis,
 and programmed cell death known as apoptosis, we also consider two states for the  ``dead'' class of subpopulations of chondrocytes.
 We note that for the purposes considered herein,  apoptotic cells do not feed back into the system.
 Due to the abrupt nature of the injury, we assume that the initial injury results in necrosis
 of cells at the injury site. Furthermore, we assume that cell death due to secondary cytokine-induced injury is strictly through apoptosis.
 The reasoning here is that necrosis is a nonspecific event that occurs in cases of severe pathological cell and tissue
damage, whereas secondary cytokine-induced injury corresponds with a physiologic form of cell death used to remove cells in a more orderly and
regulated fashion and there is evidence that often, this is via apoptosis \cite{DelCarlo2008}.

 The typical injury response can be summarized as follows. An injury results in cell necrosis and the release of alarmins (such as damage-associated molecular pattern molecules DAMPs), which initiate the chemical cascade associated with the innate
 immune and cartilage injury responses \cite{bianchi2006,harris2006}.
  The DAMPs signal healthy cells near the injury to enter the catabolic state, catabolic cells
  are capable of the production of TNF-$\alpha$ which is fundamental to inflammation.
  The inflammatory cytokine TNF-$\alpha$ has multifold effects on the system: It
  \begin{enumerate}
    \item feeds back to promote further switching of cells in the healthy state into cells in the catabolic state,
    \item causes cells in the catabolic state to enter the EPOR active state, in which they express a receptor for
      EPO and are no longer capable of synthesizing TNF-$\alpha$ \cite{brines2008},
    \item influences apoptosis of cells in the catabolic and EPOR active states,
    \item degrades extracellular matrix (denoted by $U$) which results in increased concentrations of DAMPs,
    \item has a limiting effect on production of EPO \cite{brines2008}.
  \end{enumerate}
  Catabolic cells also produce reactive oxygen species (ROS) which influences the production of EPO by healthy cells.
  We denote the concentration of ROS at a given time and location by $R$. There is a time delay of 20--24 hours before
  a healthy cell signaled by ROS will begin to produce EPO \cite{brines2008}.

  In the following we use the notation, as in \cite{graham1,graham2}, for the mathematical model of chondrocyte/cytokine interactions during injury response:
  \begin{enumerate}
    \item $R$ - concentration of reactive oxygen species (ROS) at a given time and spatial location
    \item $M$ - concentration of alarmins (DAMPs) at a given time and spatial location
    \item $F$ - concentration of the pro-inflammatory cytokine TNF-$\alpha$ at a given time and spatial location
    \item $P$ - concentration of the anti-inflammatory cytokine EPO at a given time and spatial location
    \item $U$ - density of extra-cellular matrix at a given time and spatial location
    \item $C$ - population density of healthy cells at a given time and spatial location
    \item $\sT$ - population density of catabolic cells at a given time and spatial location
    \item $\sA$ - population density of EPO receptor (EPOR) active cells at a given time and spatial location
    \item $\dN$ - population density of necrotic cells at a given time and spatial location
  \end{enumerate}

The equations making up the mathematical model developed in \cite{graham1,graham2} are
\begin{subequations}
\begin{align}
    \partial_{t}R = & \nabla\cdot(D_{R}\nabla R) - \delta_{R}R  + \sigma_{R}\sT, \label{eq:sysa2}\\
    \partial_{t}M = & \nabla\cdot(D_{M}\nabla M) - \delta_{M}M + \sigma_{M}\dN + \delta_{U}U\frac{F}{L_{F} + F},\label{eq:sysb2}\\
    \partial_{t}F  = & \nabla\cdot(D_{F}\nabla F) - \delta_{F}F + \sigma_{F}\sT, \label{eq:sysc2}\\
    \partial_{t}P  = &  \nabla\cdot(D_{P}\nabla P) - \delta_{P}P
    + \sigma_{P}C(t-\tau_{2}) \frac{R(t-\tau_{2})}{L_{R} + R(t-\tau_{2})}\frac{K_{F}}{K_{F} + F}\label{eq:sysd2}, \\
    \partial_{t} C = & \alpha\sA\frac{P}{L_{P} + P} - \beta_{1} C\frac{M}{L_{M} + M}\frac{K_{P}}{K_{P}+P} \nonumber \\
    &-\beta_{2}C\frac{F}{L_{F} + F}\frac{K_{P}}{K_{P}+P},  \label{eq:syse2}\\
    \partial_{t}\sT = & \beta_{1} C\frac{M}{L_{M} + M}\frac{K_{P}}{K_{P}+P}+\beta_{2}C\frac{F}{L_{F} + F}\frac{K_{P}}{K_{P}+P} \nonumber \\
     &- \gamma\sT(t-\tau_{1})\frac{F(t-\tau_{1})}{L_{F} + F(t-\tau_{1})} -\nu\sT\frac{F}{L_{F} + F}\frac{M}{L_{M} + M} , \label{eq:sysf2}\\
  \partial_{t}\sA = & \gamma\sT(t-\tau_{1})\frac{F(t-\tau_{1})}{L_{F} + F(t-\tau_{1})} - \alpha\sA\frac{P}{L_{P} + P}\nonumber \\ & - \mu_{\sA}\sA\frac{F}{L_{F} + F} , \label{eq:sysg2}\\
  \partial_{t}\dN = &  - \eta\dN, \label{eq:sysh2} \\
  \partial_{t}U = & -\delta_{U}U\frac{F}{L_{F} + F}. \label{eq:sysi2}
  \end{align}
  \label{eq:equations}
\end{subequations}
  Table \ref{table2} describes the meaning and units of the model parameters. The baseline parameter values for the model appear in table 1 of \cite{graham2}. By baseline we mean values that are either taken from the literature, or fit to give quantitative or qualitative agreement with biological observations.

\begin{table}
    \centering
      \begin{tabular}{|c|c|c|} \hline
        Parameter & Meaning & Units   \\ \hline
        $D_{R}$ &  Diffusion Coefficient &   $\frac{\text{cm}^{2}}{\text{day}}$  \\ \hline
        $D_{M}$ &  Diffusion Coefficient &  $\frac{\text{cm}^{2}}{\text{day}}$   \\ \hline
        $D_{F}$ &  Diffusion Coefficient &  $\frac{\text{cm}^{2}}{\text{day}}$   \\ \hline
        $D_{P}$ &  Diffusion Coefficient &  $\frac{\text{cm}^{2}}{\text{day}}$   \\ \hline
        $\delta_{R}$ & Natural Decay Rate  & $\frac{1}{\text{day}}$  \\ \hline
        $\delta_{M}$ & Natural Decay Rate  &  $\frac{1}{\text{day}}$  \\ \hline
        $\delta_{F}$ & Natural Decay Rate  & $\frac{1}{\text{day}}$  \\ \hline
        $\delta_{P}$ & Natural Decay Rate  & $\frac{1}{\text{day}}$  \\ \hline
        $\delta_{U}$ & Rate of Degradation of ECM by TNF-$\alpha$  & $\frac{1}{\text{day}}$  \\ \hline
        $\sigma_{R}$ & Production Rate  & $\frac{\text{micromolar}\cdot\text{cm}^{2}}{\text{day}\cdot\text{cells}}$  \\ \hline
        $\sigma_{M}$ & Production Rate  & $\frac{\text{micromolar}\cdot\text{cm}^{2}}{\text{day}\cdot\text{cells}}$  \\ \hline
        $\sigma_{F}$ & Production Rate  & $\frac{\text{micromolar}\cdot\text{cm}^{2}}{\text{day}\cdot\text{cells}}$  \\ \hline
        $\sigma_{P}$ & Production Rate  & $\frac{\text{micromolar}\cdot\text{cm}^{2}}{\text{day}\cdot\text{cells}}$ \\ \hline
        $K_{F}$ & Rate limiting concentration for TNF-$\alpha$ & micromolar  \\ \hline
        $K_{P}$ & Rate limiting concentration for EPO & micromolar  \\ \hline
        $L_{R}$ & Saturation constant for ROS  & micromolar  \\ \hline
        $L_{M}$ & Saturation constant for DAMPs & micromolar  \\ \hline
        $L_{F}$ & Saturation constant for TNF-$\alpha$  & micromolar  \\ \hline
        $L_{P}$ & Saturation constant for EPO  & micromolar  \\ \hline
        $\alpha$ & Response rate of EPOR active cells to EPO & $\frac{1}{\text{day}}$  \\ \hline
        $\beta_{1}$ & Response rate of healthy cells to DAMPS/EPO   & $\frac{1}{\text{day}}$  \\ \hline
        $\beta_{2}$ & Response rate of healthy cells to TNF-$\alpha$/EPO  & $\frac{1}{\text{day}}$  \\ \hline
        $\gamma$ & Response rate of catabolic cells to TNF-$\alpha$  & $\frac{1}{\text{day}}$  \\ \hline
        $\eta$ & Rate of degradation of necrotic cells & $\frac{1}{\text{day}}$  \\ \hline
        $\nu$ & Response rate of catabolic cells to TNF-$\alpha$/DAMPs  & $\frac{1}{\text{day}}$  \\ \hline
        $\mu_{S_{A}}$ & Response rate of EPOR active cells to TNF-$\alpha$  & $\frac{1}{\text{day}}$  \\ \hline
        $\tau_{1}$ & time delay in catabolic response & days  \\ \hline
        $\tau_{2}$ & time delay in production of EPO & days  \\ \hline
      \end{tabular}
      \caption{Description and units of the parameters appearing in the model (\ref{eq:sysa2})-(\ref{eq:sysi2}).}
      \label{table2}
      \end{table}

 In order to compare the simulation results with \emph{in vitro} observations it is useful to choose a ``measurable'', i.e.\ a quantity, that can
 be derived from results using the model, and easily measured from experiment. Here we consider the {\bf radius of attenuation}, this is defined to be the smallest radius beyond which a lesion cannot expand due to the actions of EPO. In the computational simulations, it is observed that the radius of attenuation varies with the change in parameter values. In the following, we will compute the radius of attenuation as certain specific parameters are varied. To remain consistent with experiment, we consider injuries to a piece of circular cartilage of diameter 2.5cm and a time frame of about ten days. In each of the results discussed below we choose a pair of parameters, then use the mathematical model to compute how the radius of attenuation varies, as the given pair of parameters is varied in a systematic way.

 How the radius of attenuation varies as dependent on a given pair of parameters tells us the influence of those parameters on the lesion expansion, or abatement during cartilage injury response. Based on this information we gain insight into which aspects of the chondrocyte/cytokine interactions are most relevant to target in potential therapies. This is one of the principal motivations for the development of the mathematical model in the first place. Furthermore, when one parameter in the given pair corresponds to a TNF-$\alpha$ related term and the other the associated EPO term, we gain insight into the details of the TNF-$\alpha$/EPO balancing act discussed in \cite{brines2008}.

\section{Results and Discussion}

  For all of the following simulations, as in \cite{graham1,graham2}, we choose the spatial domain to be a circle
  of radius 2.5 cm. This is biologically reasonable since articular cartilage is divided into three zones \cite{ulrich2003},
  with the zone forming the surface of cartilage, the superficial zone, containing the highest cell density \cite{ulrich2003}. Furthermore,  we assume circular symmetry, since the diffusion of the cytokines tend to be in the radial direction. This allows for the system (\ref{eq:equations})
  to be reduced to a problem in one spatial dimension. We choose initial conditions to represent an initial injury occurring at the
  center of the domain covering a disc of radius 0.25 cm. This is typical of the types of impact experiments that are often performed in orthopaedics labs. The boundary conditions are taken to be no-flux, i.e.\
  \begin{align}
    \left.\frac{\partial W}{\partial r}\right|_{r=2.5} = 0 \label{eq:bc}
  \end{align}
  for $W=R,M,F,P,C,\sT,\sA,\dN,U$. This essentially states that the cytokines are confined domain, and are only removed through natural decay processes.
  We note that since the system (\ref{eq:equations}) contains delay terms we must specify not only a condition at time $t=0$ but also
  a history for some time interval $(-T,0)$. For time values less than zero, the time of the initial injury, we take the history to correspond to no injury, i.e.\ the total
  cell population is in the healthy state.

  To carry out numerical approximations of the system (\ref{eq:equations}) we discretize in space as follows. Consider the diffusion equation with circular symmetry in conservative, or divergence, form
  \begin{align}
    \frac{\partial u(r,t)}{\partial t} & = \nabla_{r} \cdot J:=\frac{1}{r}\frac{\partial }{\partial r}(rJ), \label{eq:diff}
  \end{align}
   where $J = D\frac{\partial u}{\partial r}$ is the flux, and $D$ is the diffusion coefficient. Partition the radii as $r_{i}, i = 0,\ldots n$ by dividing the circle into concentric annuli. Then for $0 < i < n$ we discretize (\ref{eq:diff}) by the formula
   \begin{align}
     \pi \left(r_{i+\frac{1}{2}}^{2} - r_{i-\frac{1}{2}}^{2} \right)\frac{\partial u(r_{i},t)}{\partial t} &= 2\pi r_{i+\frac{1}{2}}J_{i+\frac{1}{2}} - 2\pi r_{i-\frac{1}{2}}J_{i-\frac{1}{2}}, \label{eq:discdiff1a}
   \end{align}
   where $J_{i\pm\frac{1}{2}}$ is the flux at $r_{i\pm\frac{1}{2}}:=\frac{r_{i}+r_{i\pm1}}{2}$, given explicitly by
   \begin{align}
     J_{i+\frac{1}{2}} &= \frac{u_{i+1} - u_{i}}{r_{i+\frac{1}{2}} - r_{i-\frac{1}{2}}} ,\\
     J_{i-\frac{1}{2}} &= \frac{u_{i} - u_{i-1}}{r_{i+\frac{1}{2}} - r_{i-\frac{1}{2}}} .
   \end{align} 
   This leads to
   \begin{align}
   \frac{\partial u(r_{i},t)}{\partial t}& = \Delta_{i}:=\frac{r_{i+\frac{1}{2}}J_{i+\frac{1}{2}} - r_{i-\frac{1}{2}}J_{i-\frac{1}{2}}}{\frac{1}{2}\left(r_{i+\frac{1}{2}}^{2} - r_{i-\frac{1}{2}}^{2} \right)}. \label{eq:discdiff1b}
   \end{align}
   We observe that 
   \begin{align}
   \frac{1}{2}\left(r_{i+\frac{1}{2}}^{2} - r_{i-\frac{1}{2}}^{2} \right) &= \frac{1}{2}(r_{i+\frac{1}{2}} + r_{i-\frac{1}{2}})(r_{i+\frac{1}{2}} - r_{i-\frac{1}{2}}) \\
   & = r_{i}\delta r_{i}, 
   \end{align}
   where $r_{i} = \frac{r_{i+\frac{1}{2}}+r_{i-\frac{1}{2}}}{2}$, and $\delta r_{i} = r_{i+\frac{1}{2}} - r_{i-\frac{1}{2}}$. Thus, the scheme (\ref{eq:discdiff1b}) corresponds to the standard finite difference approximation in polar coordinates, see for example \cite{thomas1995}. 

   For the case $i=0$, that is, at the center of the circle, we have
     \begin{align}
       \pi r_{\frac{1}{2}}^2 \frac{\partial u(0,t)}{\partial t}& = 2\pi r_{\frac{1}{2}}J_{\frac{1}{2}}, \label{eq:discdiff2a}
     \end{align}
     which gives
     \begin{align}
        \frac{\partial u(0,t)}{\partial t} &=  \Delta_{0} := \frac{J_{\frac{1}{2}}}{\frac{1}{2}r_{\frac{1}{2}}}. \label{eq:discdiff2b}
     \end{align}

     Finally, for a no-flux boundary condition as in (\ref{eq:bc}), the differencing is given by
     \begin{align}
       \pi (r_{n}^2 - r_{n-\frac{1}{2}}^2) \frac{\partial u(r_{n},t)}{\partial t}& = -2\pi r_{n-\frac{1}{2}}J_{n-\frac{1}{2}}, \label{eq:discdiff3a}
     \end{align}
     which gives
     \begin{align}
       \frac{\partial u(r_{n},t)}{\partial t} &= \Delta_{n}:=\frac{-r_{n-\frac{1}{2}}J_{n-\frac{1}{2}}}{\frac{1}{2}(r_{n}^2 - r_{n-\frac{1}{2}}^2)}. \label{eq:discdiff3b}
     \end{align}
   These formulas are reproduced from appendix C of \cite{bruceThesis}\footnote{We note that in \cite{bruceThesis} there are misprints in the formulas corresponding to (\ref{eq:discdiff3a}), (\ref{eq:discdiff3b}) which have here been corrected.}

   Applying (\ref{eq:discdiff1b}), (\ref{eq:discdiff2b}), and (\ref{eq:discdiff3b}) to the spatial terms in (\ref{eq:sysa2}),(\ref{eq:sysb2}),(\ref{eq:sysc2}), and (\ref{eq:sysd2}) then gives a semi-discrete system of delay-differential equations
 \begin{subequations}
   \begin{align}
    \partial_{t}R_{i} = & \Delta_{i} - \delta_{R}R_{i}  + \sigma_{R}(\sT)_{i}, \label{eq:sysa3}\\
    \partial_{t}M_{i} = & \Delta_{i} - \delta_{M}M_{i} + \sigma_{M}(\dN)_{i} + \delta_{U}U_{i}\frac{F_{i}}{L_{F} + F_{i}},\label{eq:sysb3}\\
    \partial_{t}F{i}  = & \Delta_{i} - \delta_{F}F_{i} + \sigma_{F}(\sT)_{i}, \label{eq:sysc3}\\
    \partial_{t}P_{i}  = &  \Delta_{i} - \delta_{P}P_{i}
    + \sigma_{P}C_{i}(t-\tau_{2}) \frac{R_{i}(t-\tau_{2})}{L_{R} + R_{i}(t-\tau_{2})}\frac{K_{F}}{K_{F} + F_{i}}\label{eq:sysd3}, \\
    \partial_{t} C_{i} = & \alpha(\sA)_{i}\frac{P_{i}}{L_{P} + P_{i}} - \beta_{1} C_{i}\frac{M_{i}}{L_{M} + M_{i}}\frac{K_{P}}{K_{P}+P_{i}} \nonumber \\
    &-\beta_{2}C_{i}\frac{F_{i}}{L_{F} + F_{i}}\frac{K_{P}}{K_{P}+P_{i}},  \label{eq:syse3}\\
    \partial_{t}(\sT)_{i} = & \beta_{1} C_{i}\frac{M_{i}}{L_{M} + M_{i}}\frac{K_{P}}{K_{P}+P_{i}}+\beta_{2}C_{i}\frac{F_{i}}{L_{F} + F_{i}}\frac{K_{P}}{K_{P}+P_{i}} \nonumber \\
     &- \gamma(\sT)_{i}(t-\tau_{1})\frac{F_{i}(t-\tau_{1})}{L_{F} + F_{i}(t-\tau_{1})} -\nu(\sT)_{i}\frac{F_{i}}{L_{F} + F_{i}}\frac{M_{i}}{L_{M} + M_{i}} , \label{eq:sysf3}\\
  \partial_{t}(\sA)_{i} = & \gamma(\sT)_{i}(t-\tau_{1})\frac{F_{i}(t-\tau_{1})}{L_{F} + F_{i}(t-\tau_{1})} - \alpha(\sA)_{i}\frac{P_{i}}{L_{P} + P_{i}}\nonumber \\ & - \mu_{\sA}(\sA)_{i}\frac{F_{i}}{L_{F} + F_{i}} , \label{eq:sysg3}\\
  \partial_{t}(\dN)_{i} = &  - \eta(\dN)_{i}, \label{eq:sysh3} \\
  \partial_{t}U_{i} = & -\delta_{U}U_{i}\frac{F_{i}}{L_{F} + F_{i}}, \label{eq:sysi3}
  \end{align}
  \label{eq:discreteEquations}
\end{subequations}
   for $i=0,\ldots, n$, where $\Delta_{i}$ is the appropriate discrete circularly symmetric diffusion operator from (\ref{eq:discdiff1b}),(\ref{eq:discdiff2b}), or (\ref{eq:discdiff3b}).
  The semi-discrete system is solved in MATLAB using the delay-differential equation solver {\tt dde23}.
  For details on the methods and software for solving delay-differential equations see \cite{larry1,larry2,larry3}.

  The primary parameter pairs of interest are the time delays $\tau_{1},\tau_{2}$, the diffusion coefficients $D_{F},D_{P}$ for TNF-$\alpha$ and EPO respectively, the production rates $\sigma_{F},\sigma_{P}$ for TNF-$\alpha$ and EPO respectively, and the saturation constants $K_{F},K_{P}$ for TNF-$\alpha$ and EPO respectively. These are the parameters that are most closely tied to the balancing act between pro- and anti-inflammatory
  cytokines, and this is what is of primary interest to researchers working to develop therapies to minimize the collateral damage associated with inflammation in cartilage injuries.

\begin{figure}[hbtp]
  \centering
  \includegraphics[width=75mm,height=75mm]{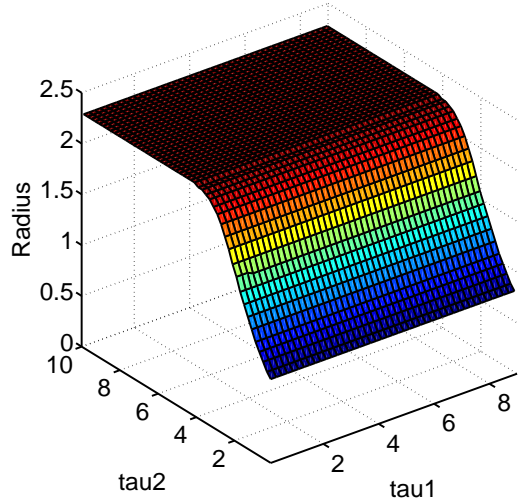}\\
  \caption{Radius of attenuation as it varies simultaneously with $\tau_{1},\tau_{2}$, with other parameters held fixed. }\label{delayVary1}
\end{figure}

  The first pair of parameters we vary are the time delay parameters $\tau_{1},\tau_{2}$. We recall that $\tau_{1}$ is the delay that for catabolic cells signaled by TNF-$\alpha$ to become EPOR active, while $\tau_{2}$ is the delay for a healthy cell signaled by reactive oxygen species (ROS) to synthesize EPO. The baseline values for the delays are 12 hours for $\tau_{1}$ and 24 hours for $\tau_{2}$ \cite{brines2008,graham2}. Figure \ref{delayVary1} shows the radius of attenuation as it varies with $\tau_{1},\tau_{2}$ over the domain $[0,10]\times [0,10]$ with units in days.
  We observe that the delay parameter $\tau_{2}$ has a more significant impact on the radius on attenuation than does $\tau_{1}$. Since $\tau_{2}$
  corresponds to the delay in a healthy cell signaled by reactive oxygen species to produce EPO, our results support the hypothesis in \cite{brines2008} that intervention with exogenous EPO is an important step in limiting the amount of collateral damage caused by TNF-$\alpha$.

\begin{figure}[hbtp]
  \centering
  \includegraphics[width=75mm,height=75mm]{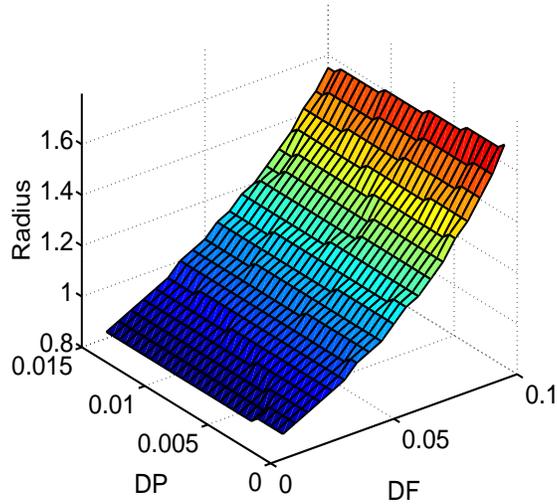}\\
  \caption{Radius of attenuation as it varies simultaneously with $D_{F},D_{P}$, with other parameters held fixed.  }\label{diffVary2}
\end{figure}

Now we consider the effects of varying the diffusion coefficients $D_{F},D_{P}$ for TNF-$\alpha$ and EPO respectively. We note that the
diffusion of cytokines is the principal mechanism that determines the spatial behavior of lesion spreading in articular cartilage. The baseline
values for $D_{F},D_{P}$ are 0.05, 0.005 $\frac{\text{mm}^{2}}{\text{day}}$ respectively \cite{graham2,Leddy2003}. Figure \ref{diffVary2}
shows the radius of attenuation as a function of $D_{F},D_{P}$ over the domain $[0,0.1]\times[0,0.015]$. We observe that, of the two diffusion parameters, $D_{F}$ has the greater impact on the radius of attenuation. This is somewhat expected in light of the fact that
there is a time delay for production of EPO by healthy cells signaled by ROS. Because of this delay TNF-$\alpha$ is typically produced at significantly earlier times than EPO. Thus the degree to which TNF-$\alpha$ can diffuse significantly influences how far the lesion can spread during this initial time period before there are sufficient concentrations of EPO to abate the spread of damage.

\begin{figure}[hbtp]
  \centering
  \includegraphics[width=75mm,height=75mm]{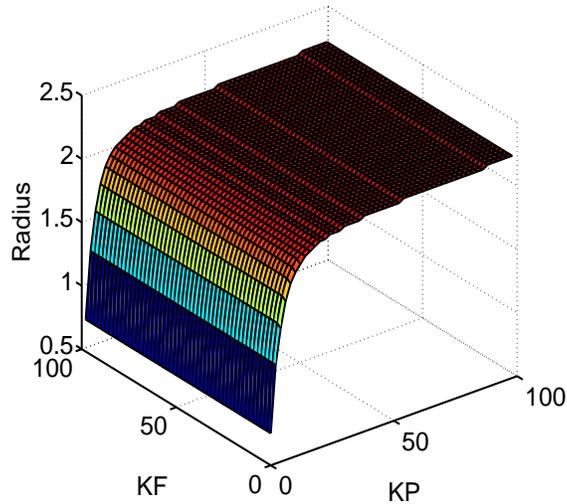}\\
  \caption{Radius of attenuation as it varies simultaneously with $K_{F},K_{P}$, with other parameters held fixed. }\label{satVary1}
\end{figure}

\begin{figure}[hbtp]
  \centering
  \includegraphics[width=75mm,height=75mm]{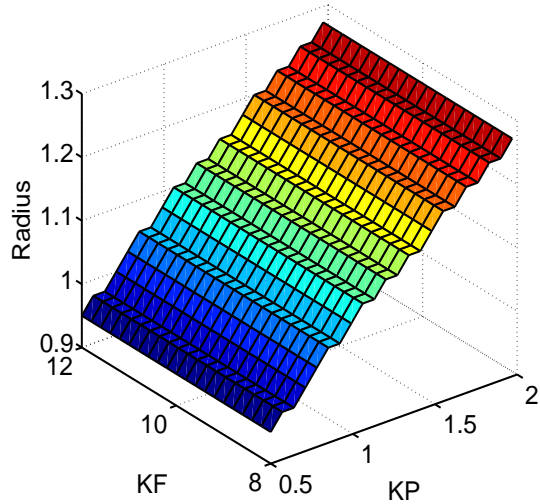}\\
  \caption{ Radius of attenuation as it varies simultaneously with $K_{F},K_{P}$, with other parameters held fixed. }\label{satVary2}
\end{figure}

Next, we consider the radius of attenuation as it varies simultaneously with the saturation constants $K_{F}$ and $K_{P}$. The baseline values for these parameters are 10 and 1 respectively, with units in micromoles. Figure \ref{satVary1} shows the radius of attenuation as a function of $K_{F},K_{P}$ over the domain $[0,100]\times[0,100]$ with units in micromoles. This is quite a large variation for $K_{F}$ and $K_{P}$. We observe that the saturation constant, $K_{P}$, for EPO to limit the response of healthy cells to TNF-$\alpha$ and alarmins (DAMPs) has the greater influence of the two parameters $K_{F},K_{P}$ on the radius of attenuation. Thus the results of the mathematical model seem to suggest that healthy cells must be sensitive to relatively low concentrations of EPO in order to minimize damage, and for maximal healing to occur. Figure \ref{satVary2} again shows the radius of attenuation as a function of $K_{F},K_{P}$ but for a smaller range in the parameter values. Here we focus on values relatively close to the baseline values, this gives an idea of how  sensitive the model is to small, simultaneous changes in values for the parameters $K_{F},K_{P}$.  The results are consistent with those shown in figure \ref{satVary1}.

\begin{figure}[hbtp]
  \centering
  \includegraphics[width=75mm,height=75mm]{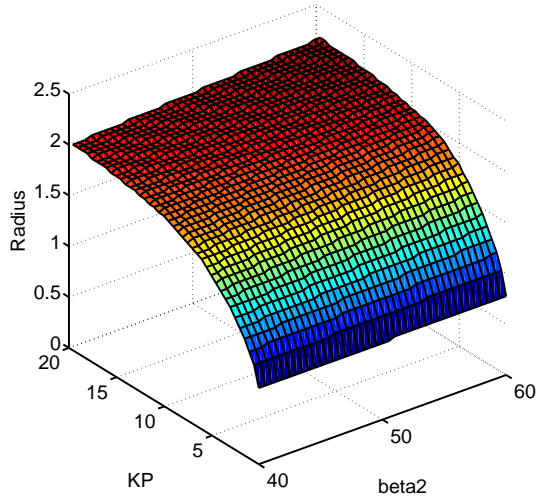}\\
  \caption{ Radius of attenuation as it varies simultaneously with $\beta_{2},K_{P}$, with other parameters held. }\label{betaKPVary1}
\end{figure}

We observed that when comparing the saturation parameters $K_{F},K_{P}$, the parameter $K_{P}$ has the more significant effect on the radius of
attenuation. However, the term involving $K_{P}$ and TNF-$\alpha$ in the model system (\ref{eq:sysa2})-(\ref{eq:sysi2}) is
\begin{equation}
  \beta_{2}C\frac{F}{L_{F} + F}\frac{K_{P}}{K_{P}+P},
\end{equation}
which influences the switch from the healthy to the catabolic state. Thus, the switching is determined by the parameters $\beta_{2}$ and $K_{P}$
together. We examine the radius of attenuation as a function of $\beta_{2},K_{P}$, the results are shown in figure \ref{betaKPVary1}. Again it is observed that $K_{P}$ has the more significant impact in determining the radius of attenuation.

\begin{figure}[hbtp]
  \centering
  \includegraphics[width=75mm,height=75mm]{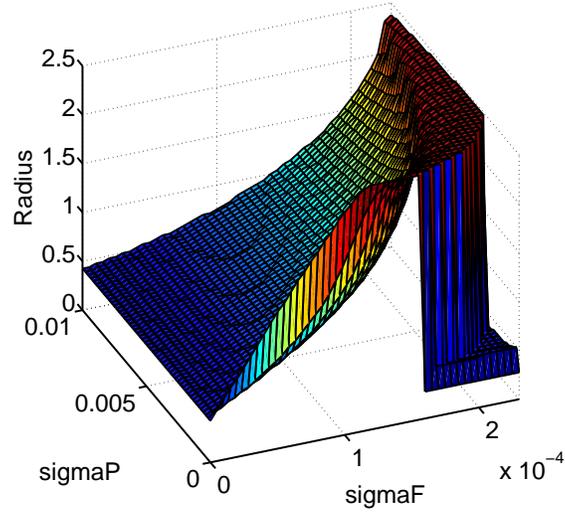}\\
  \caption{ Radius of attenuation as it varies simultaneously with $\sigma_{F},\sigma_{P}$, with other parameters held. }\label{sigmaVary1}
\end{figure}

\begin{figure}[hbtp]
  \centering
  \includegraphics[width=75mm,height=75mm]{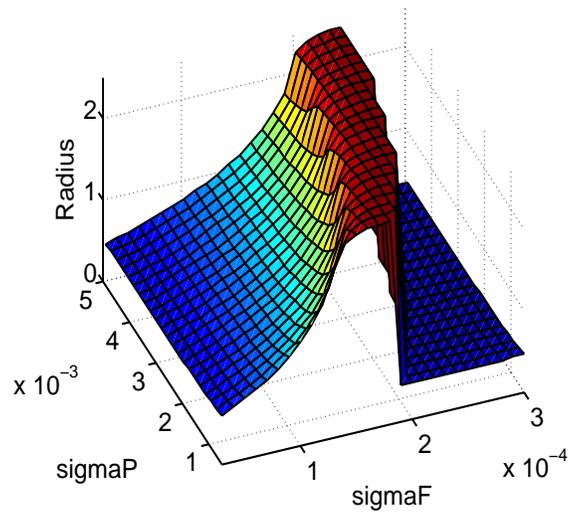}\\
  \caption{Radius of attenuation as it varies simultaneously with $\sigma_{F},\sigma_{P}$, with other parameters held fixed. }\label{sigmaVary3}
\end{figure}

We next examine the radius of attenuation as as it varies simultaneously with the production rates $\sigma_{F},\sigma_{P}$ of TNf-$\alpha$ and EPO respectively. The baseline values for these parameters are 0.0001 for $\sigma_{F}$ and 0.001 for $\sigma_{P}$ with units in $\text{days}^{-1}$. Figure \ref{sigmaVary1} shows the radius of attenuation as a function of $\sigma_{F},\sigma_{P}$ over the domain $[0,0.0001]\times[0,0.1]$. We observe that, overall, the radius of attenuation increases as $\sigma_{F}$, the production of TNF-$\alpha$, increases. However, the radius of attenuation appears to be a nonlinear function of $\sigma_{F},\sigma_{P}$. In figure \ref{sigmaVary3} we show the radius of attenuation as a function of $\sigma_{F},\sigma_{P}$ over the domain $[0,0.0003]\times[0.001,0.005]$. This shows that, while the radius of attenuation is generally increasing
as a function of $\sigma_{F}$, that for a fixed value of $\sigma_{P}$ there is a point beyond which the radius of attenuation exceeds the domain. Thus, for our domain, if $\sigma_{F}$ is sufficiently large then there is no radius of attenuation. However, this does not necessarily imply
that no healing whatsoever can take place.

\begin{figure}[hbtp]
   \centering
   \subfigure[t=0]{\includegraphics[height=1.7in]{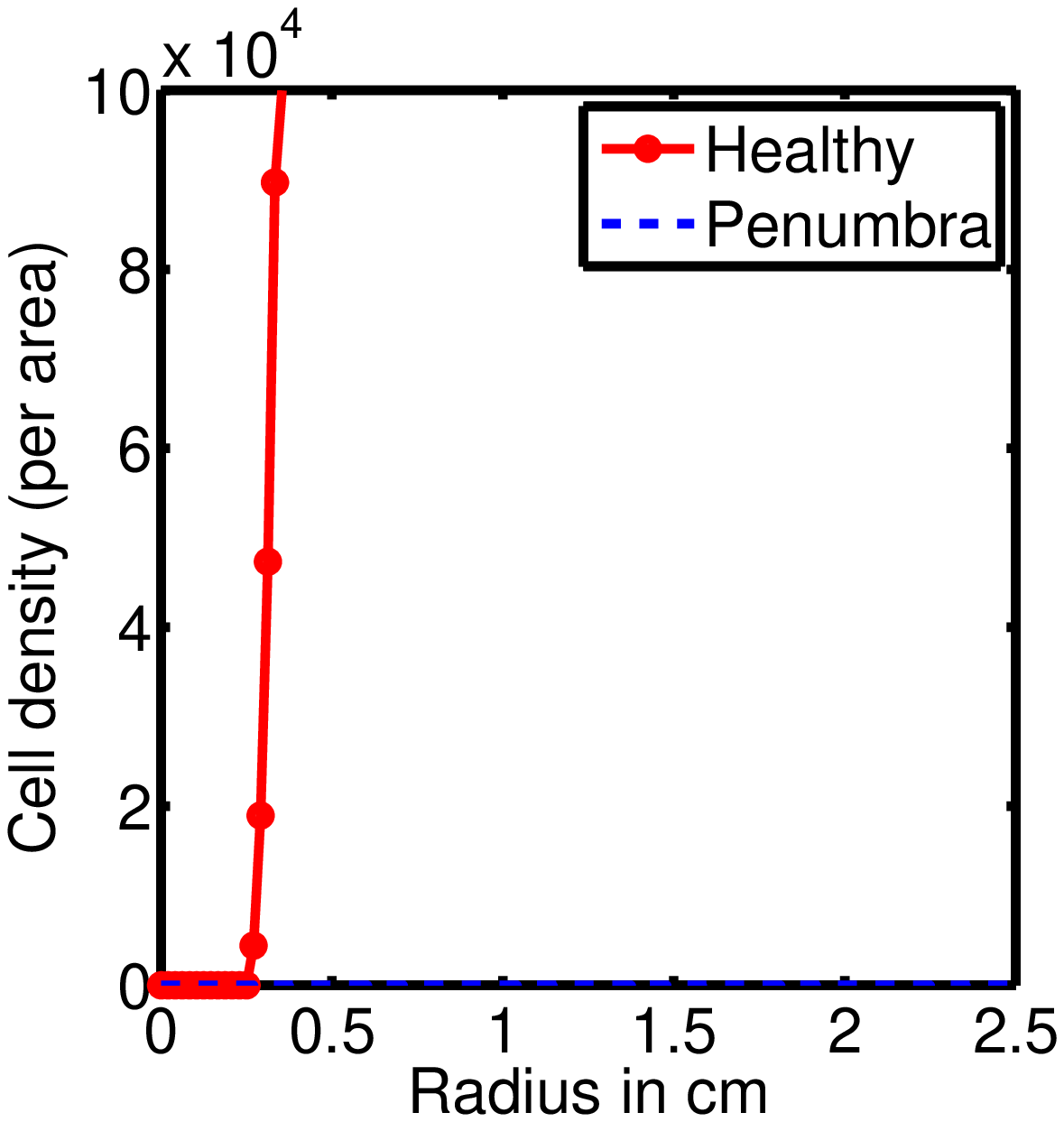}\label{dee-a}}
   \subfigure[t=1.25 days]{\includegraphics[height=1.7in]{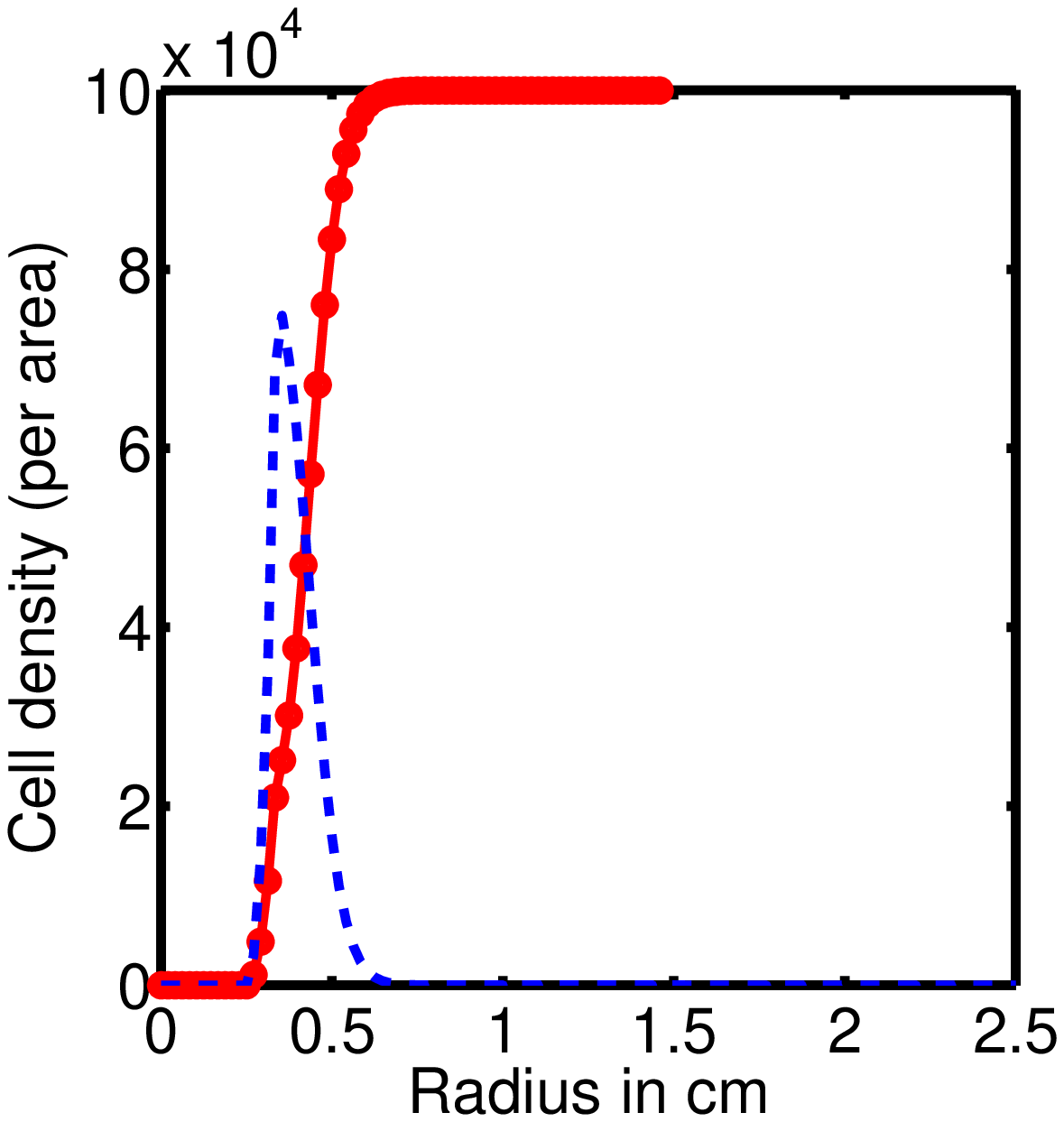}\label{dee-b}}
   \subfigure[t=2.5 days]{\includegraphics[height=1.7in]{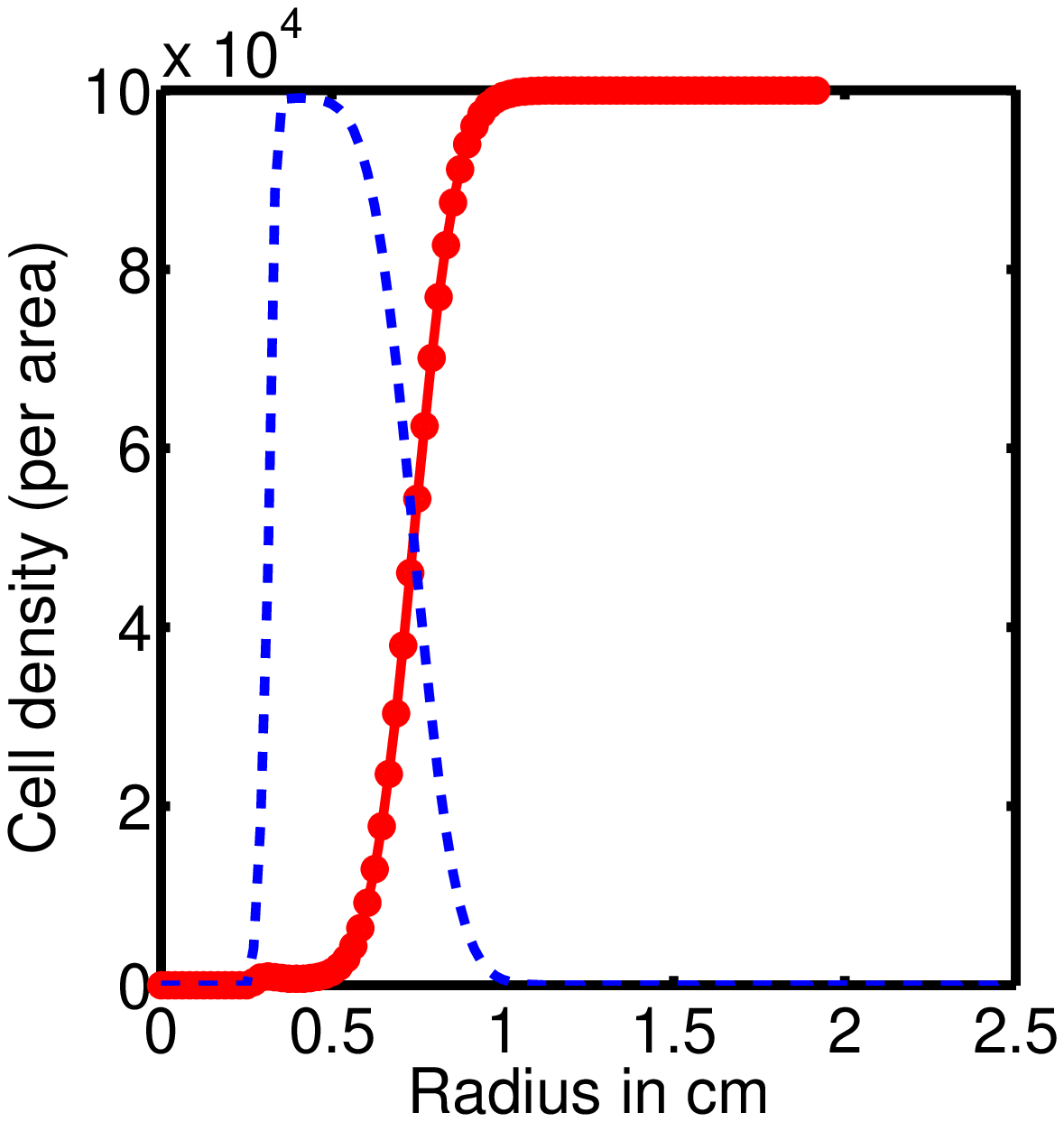}\label{dee-c}}\\
   \subfigure[t=3.75 days]{\includegraphics[height=1.7in]{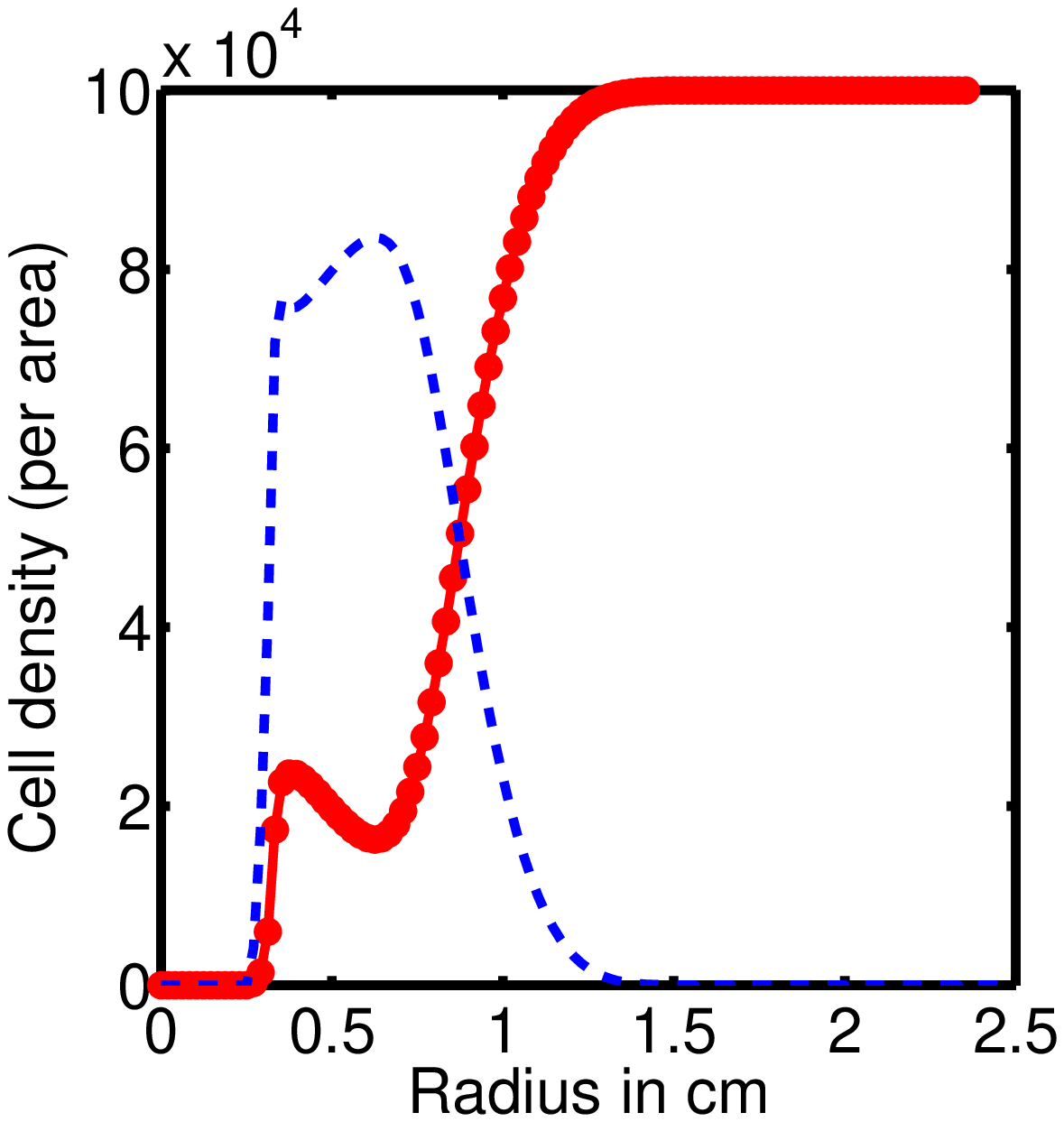}\label{dee-d}}
   \subfigure[t=5 days]{\includegraphics[height=1.7in]{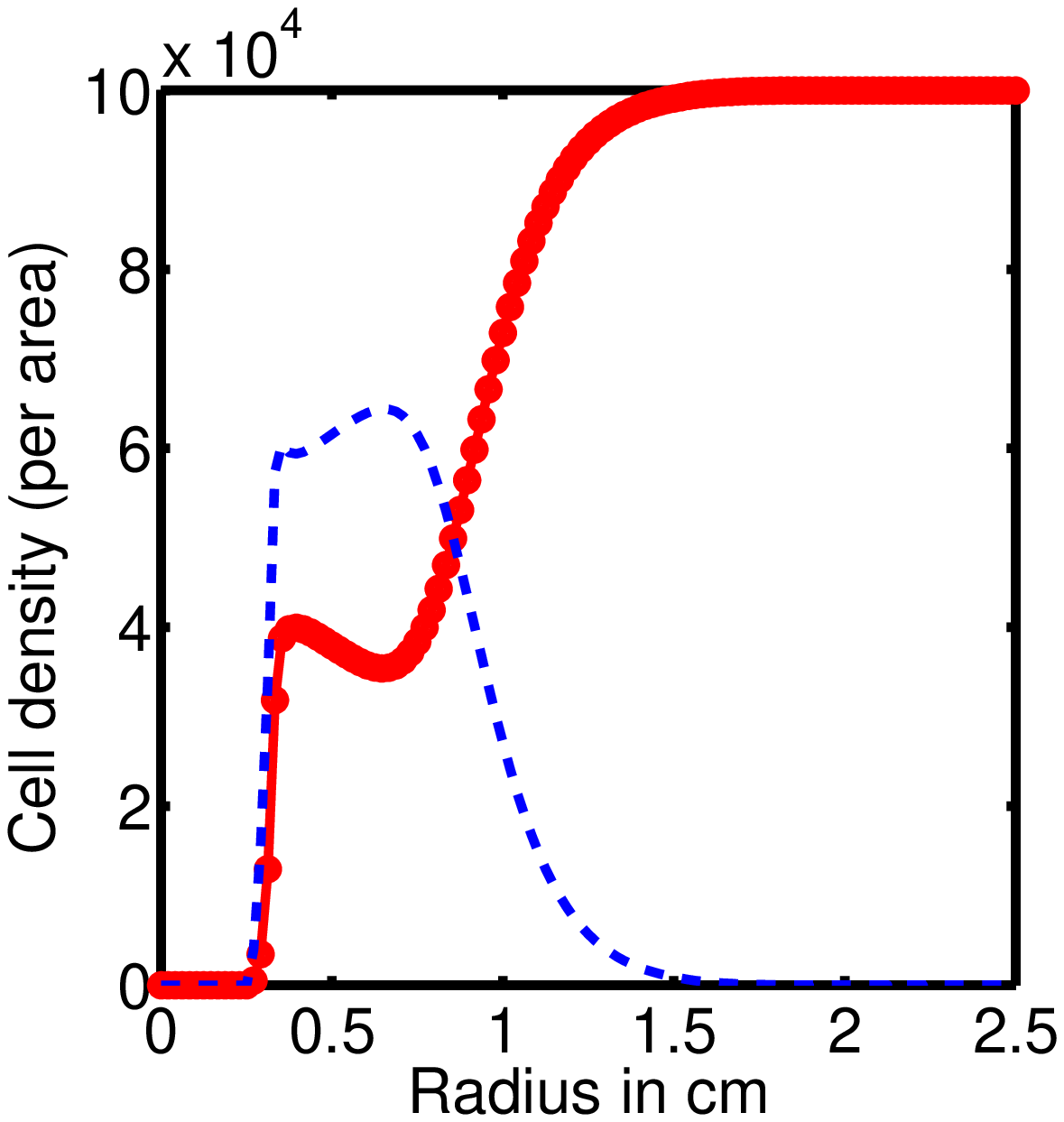}\label{dee-e}}
   \subfigure[t=6.25 days]{\includegraphics[height=1.7in]{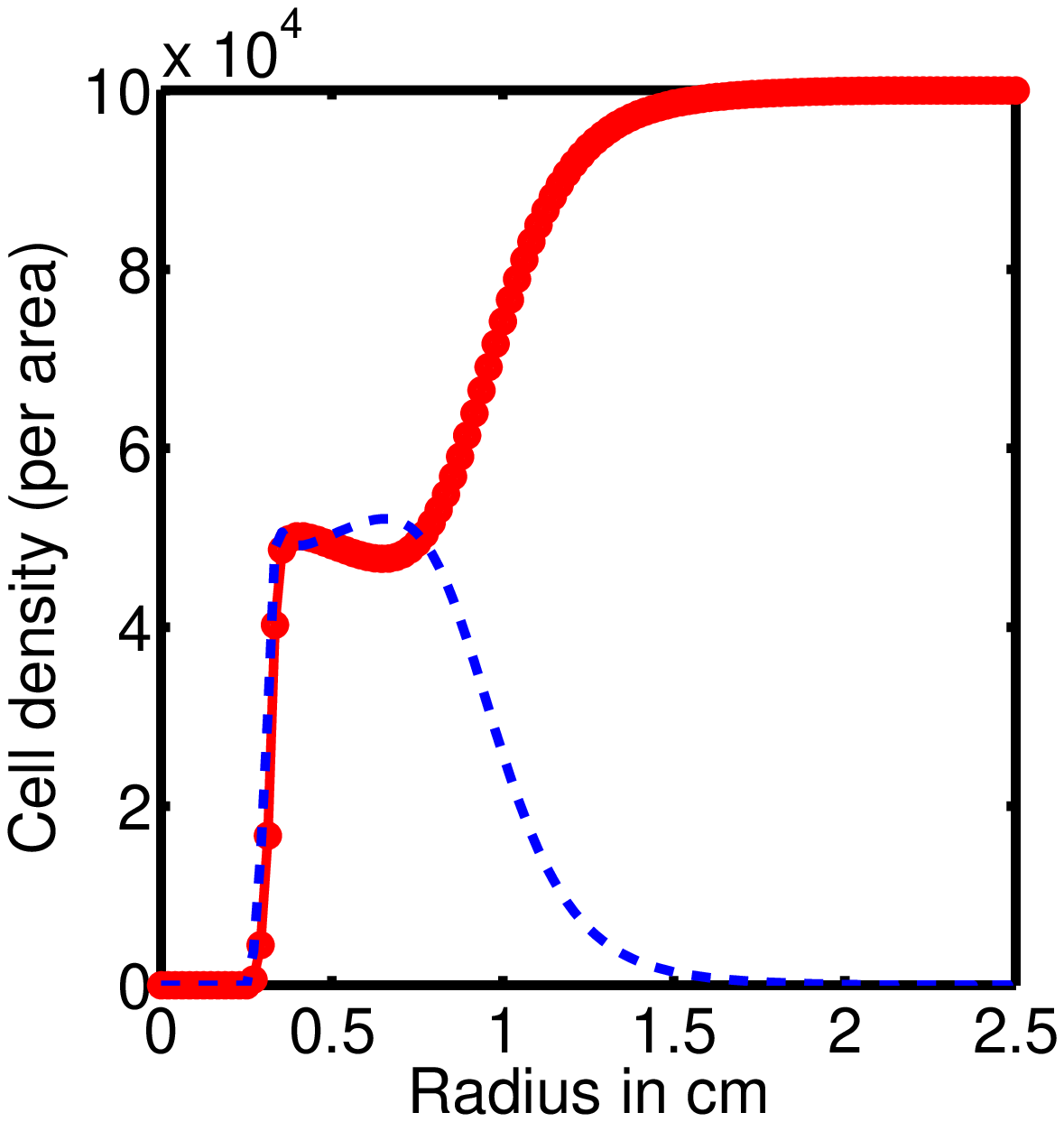}\label{dee-f}} \\
   \subfigure[t=7.5 days]{\includegraphics[height=1.7in]{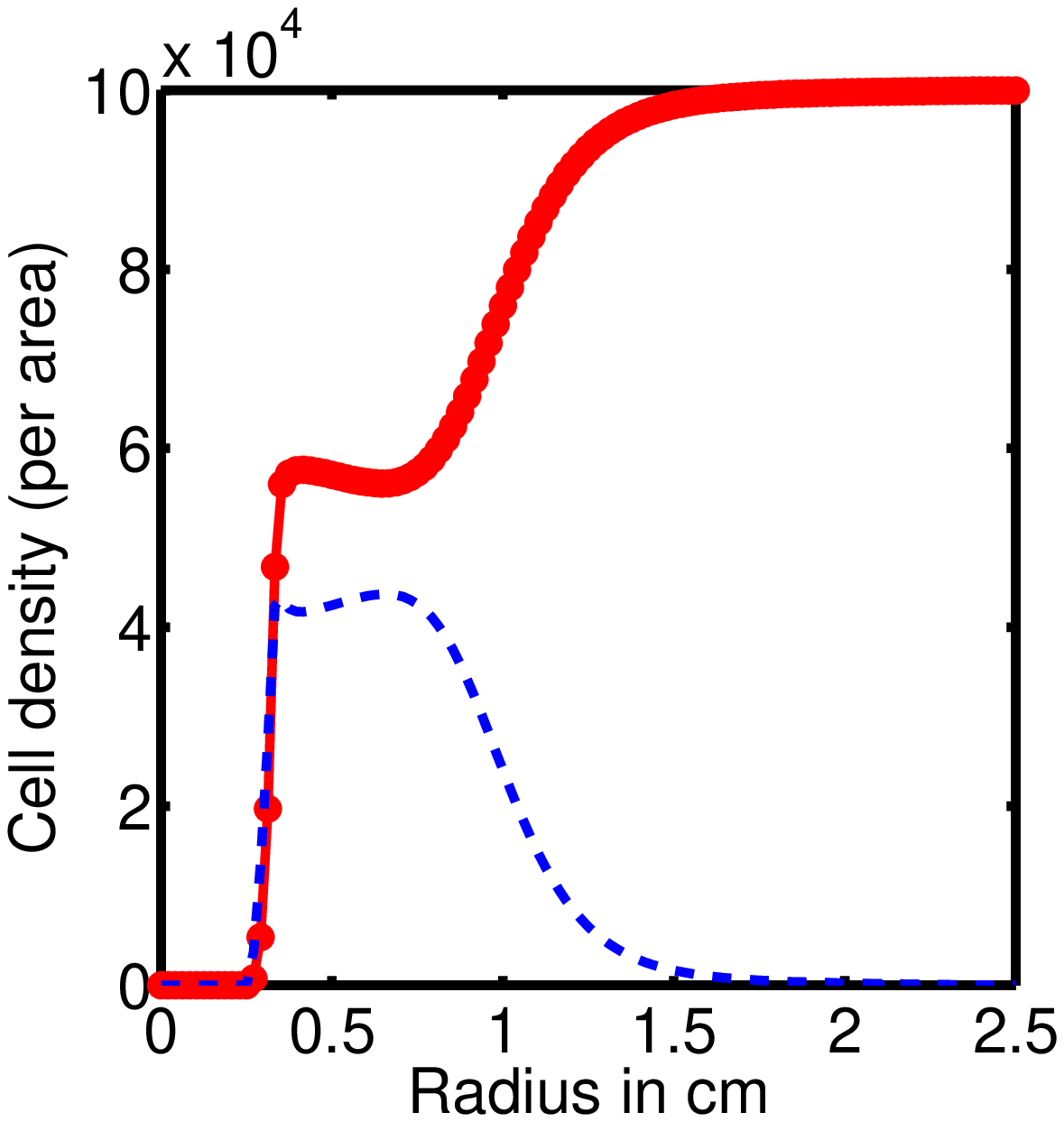}\label{dee-g}}
   \subfigure[t=10 days]{\includegraphics[height=1.7in]{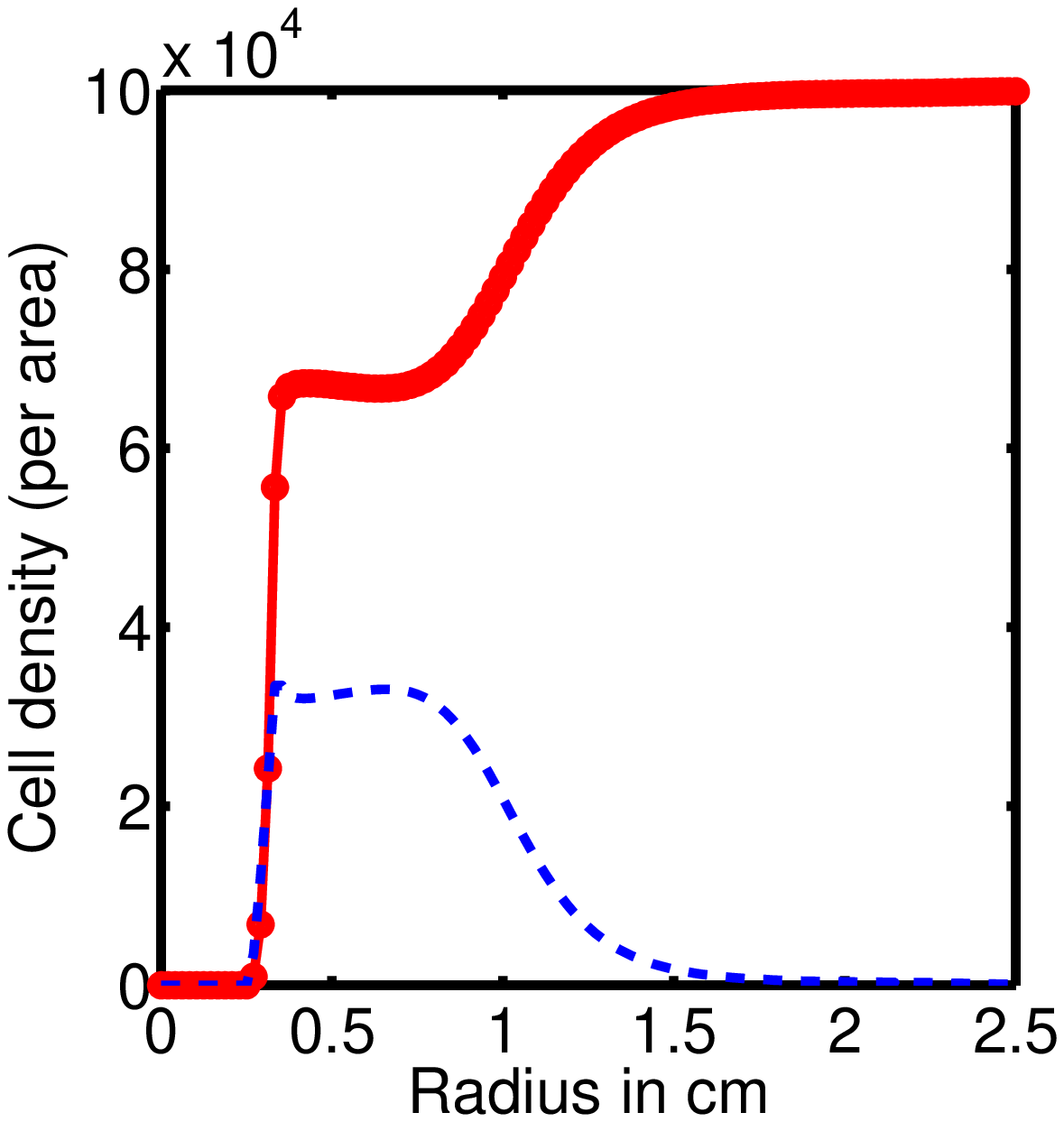}\label{dee-h}}
   \subfigure[t=20]{\includegraphics[height=1.7in]{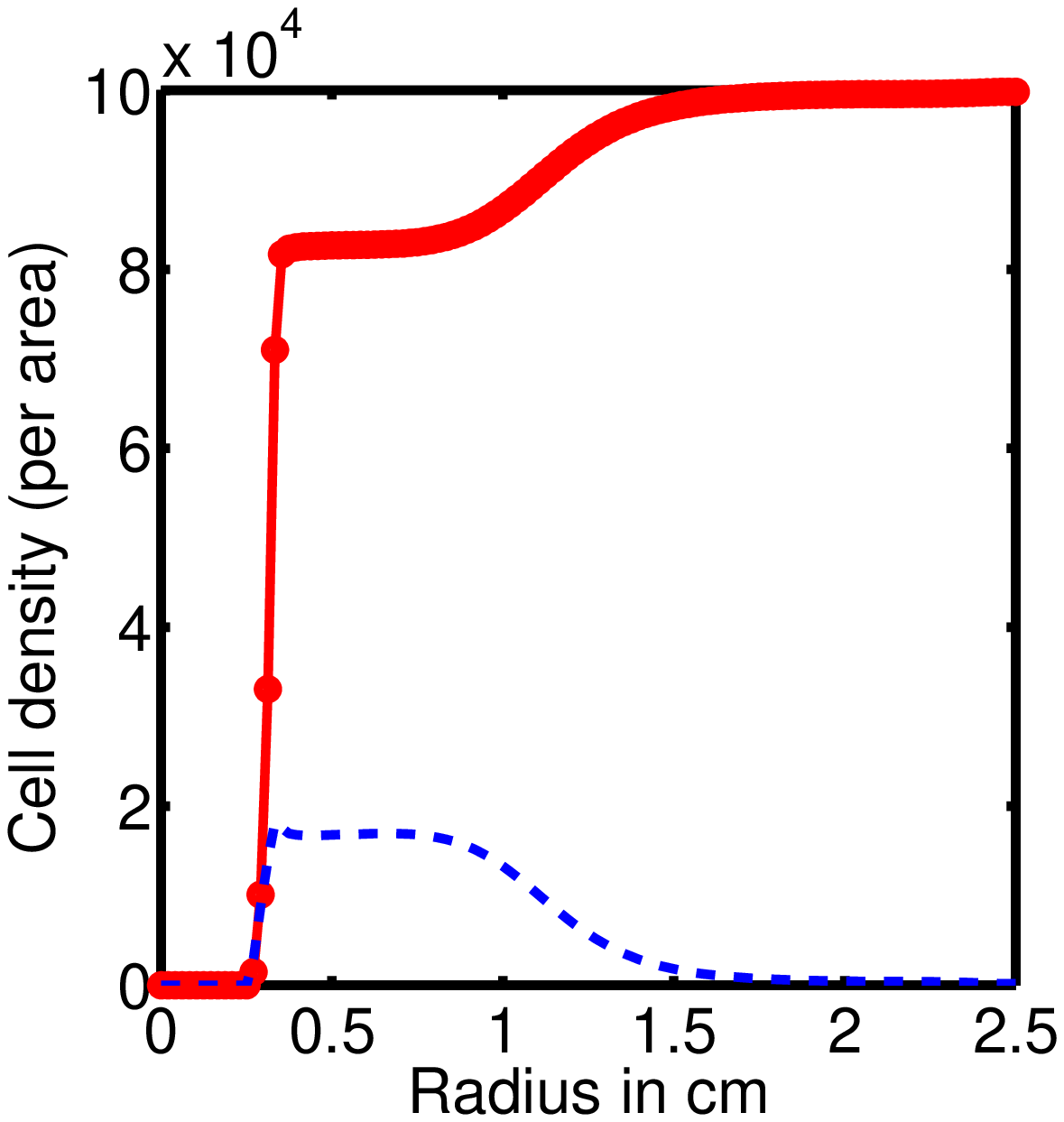}\label{dee-i}}
   \caption{Cell populations when $\sigma_{F}=0.00017$ and $\sigma_{P}=0.0032$.}
   \label{wtf1}
 \end{figure}

Figure \ref{wtf1} shows the dynamics of the healthy and penumbral (sum of catabolic and EPOR active) cell populations for values
of $\sigma_{F}$ and $\sigma_{P}$ that, according to figure \ref{sigmaVary3}, lead to a radius of attenuation of approximately 1.7cm.
Since the radius of attenuation in figure \ref{sigmaVary3} is computed based on a ten day time period, figure \ref{dee-i} shows the
cell populations as a function of radius after a twenty day period to ensure that the radius does not continue to expand after ten days.
We observe that, while the radius of attenuation is larger than that in figure 6 of \cite{graham2}, there is still at least as much healing near the initial injury site due to EPOR active cells switching back to healthy as a result of EPO signaling. An interesting observation is the ``dip'' in the healthy cell population between radius $r=0.25cm$ and $r=0.7$ if figure \ref{dee-d}. This is due to the fact that diffusion of TNf-$\alpha$
results in lower concentrations of TNF-$\alpha$ for smaller radius values where there is
a higher concentration of EPOR active cells. Thus, the EPOR active cells can more effectively response to EPO.

\begin{figure}[hbtp]
   \centering
   \subfigure[t=0]{\includegraphics[height=1.7in]{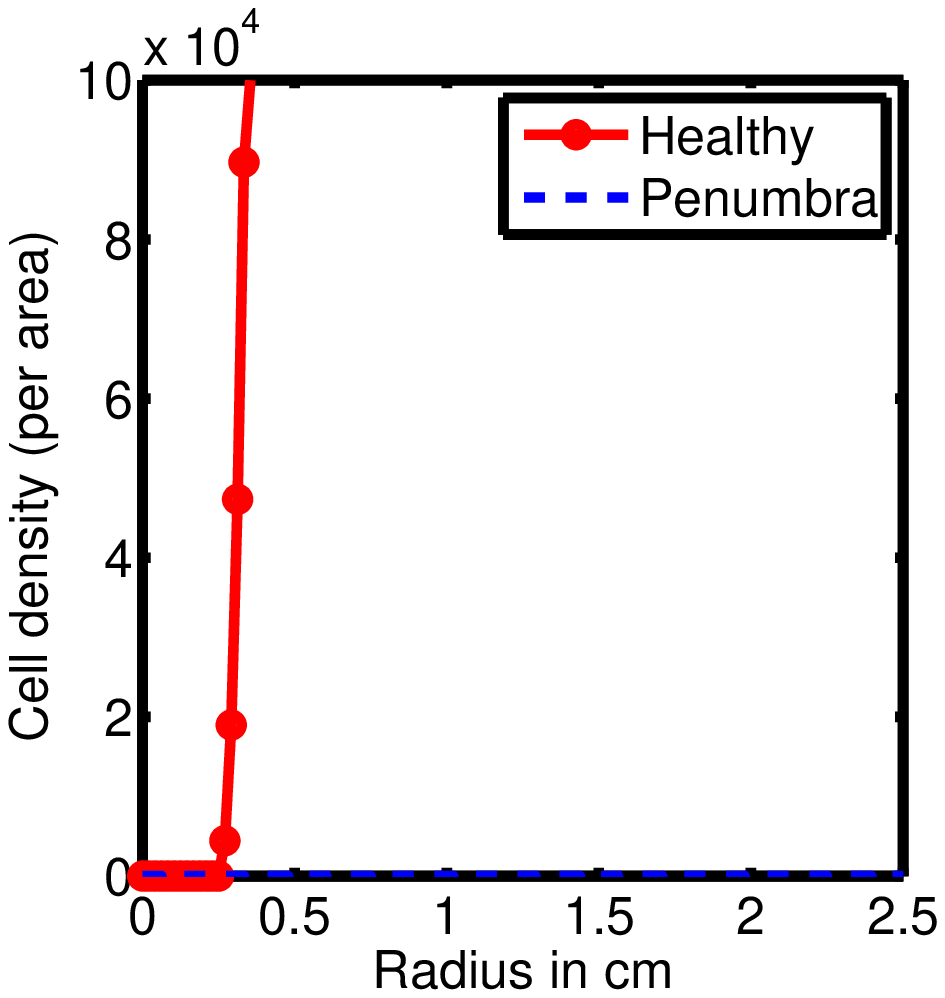}\label{dep-a}}
   \subfigure[t=1.25 days]{\includegraphics[height=1.7in]{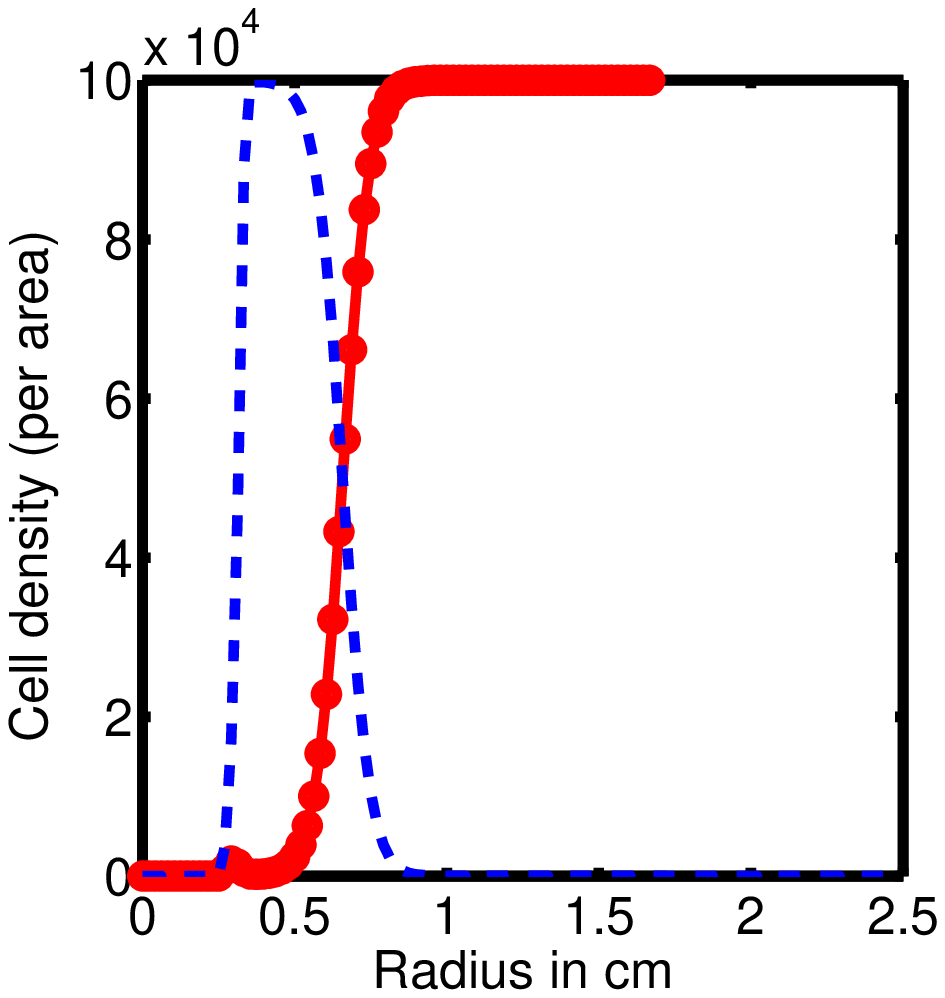}\label{dep-b}}
   \subfigure[t=2.5 days]{\includegraphics[height=1.7in]{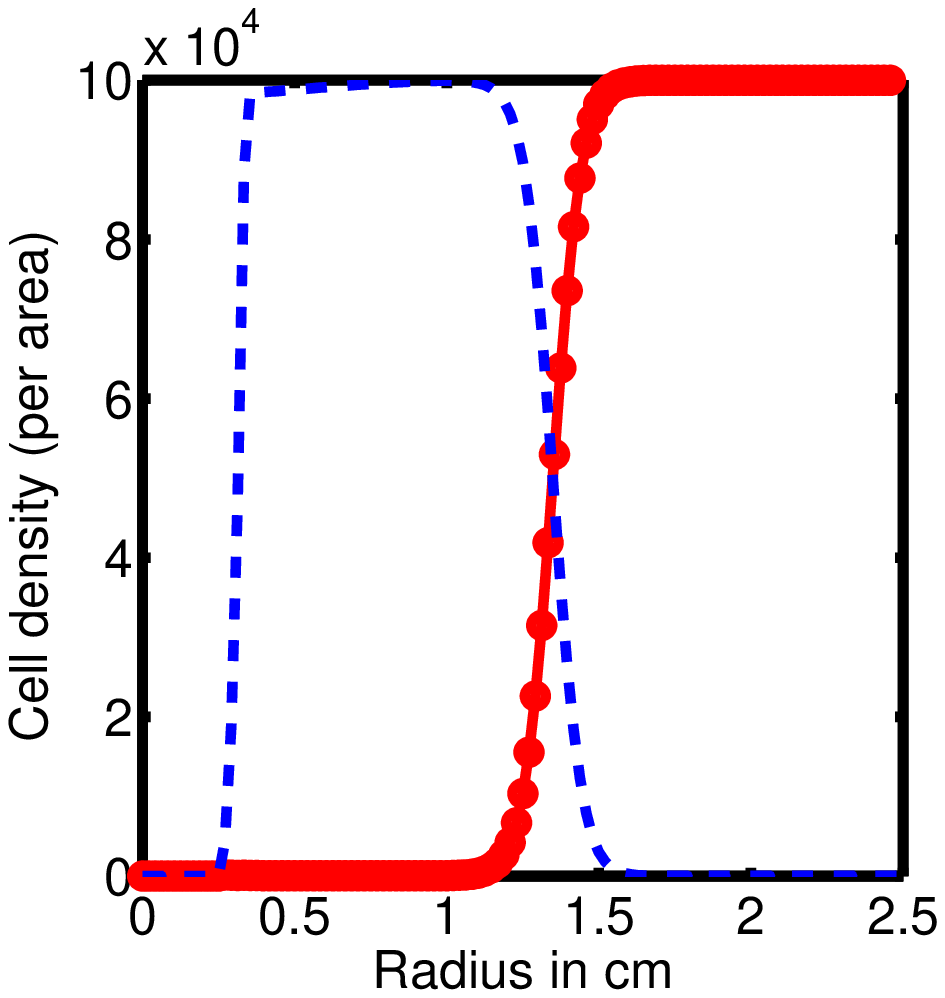}\label{dep-c}}\\
   \subfigure[t=3.75 days]{\includegraphics[height=1.7in]{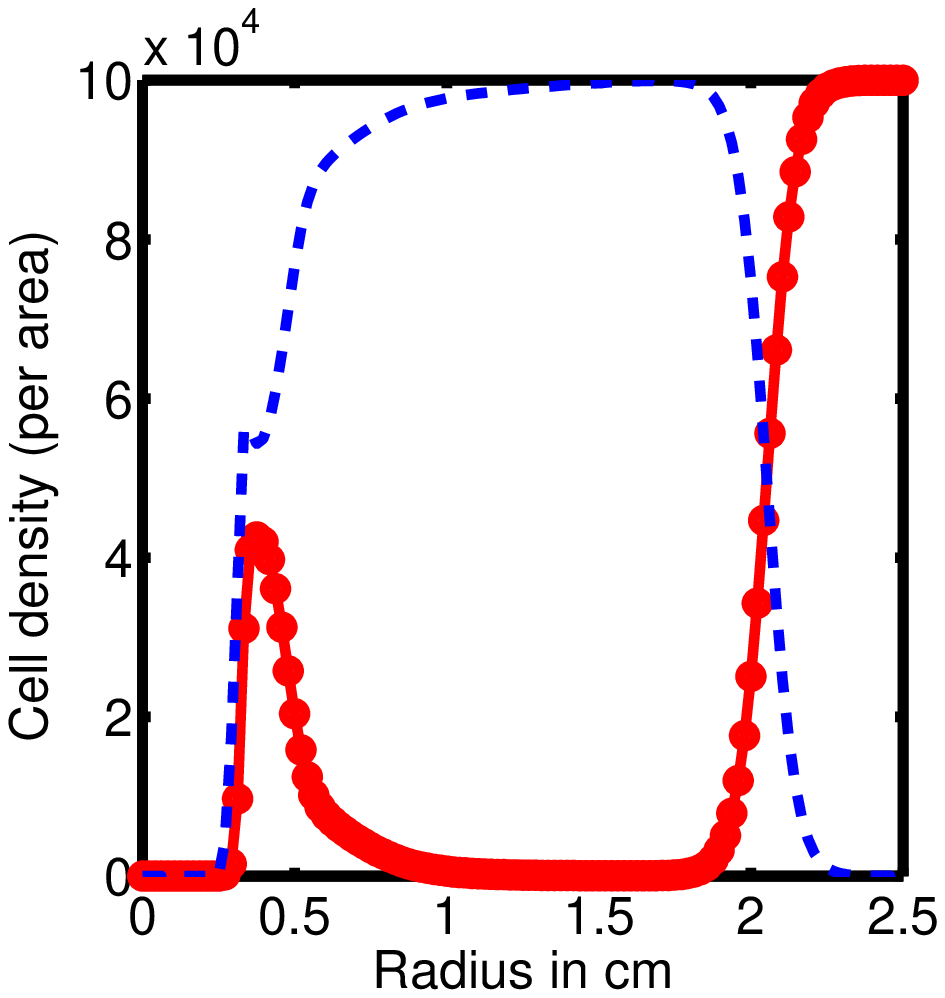}\label{dep-d}}
   \subfigure[t=5 days]{\includegraphics[height=1.7in]{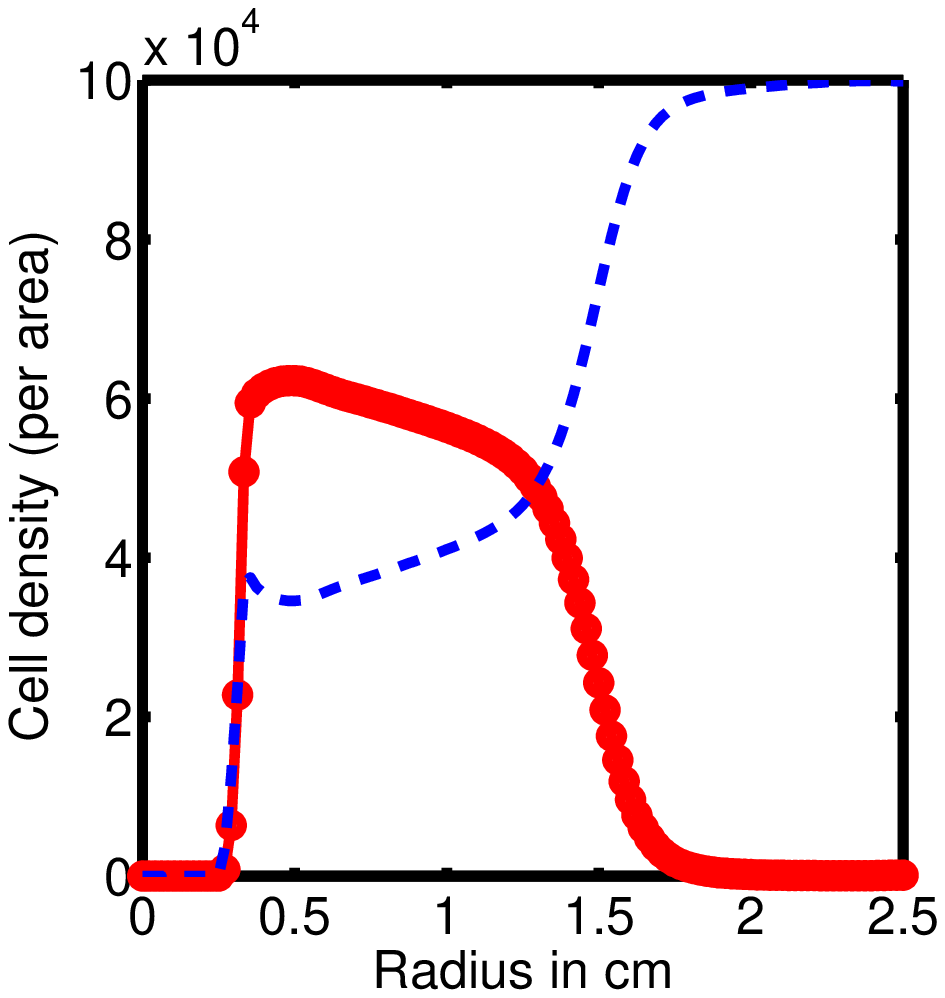}\label{dep-e}}
   \subfigure[t=6.25 days]{\includegraphics[height=1.7in]{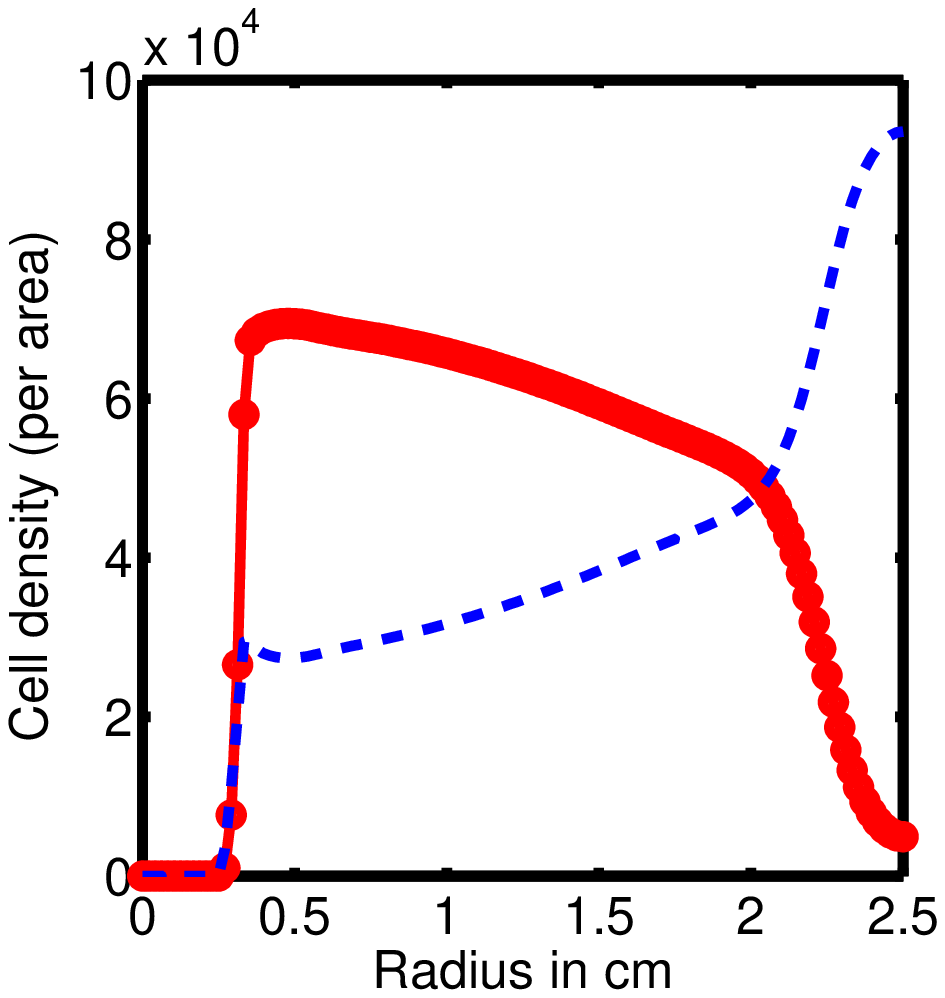}\label{dep-f}} \\
   \subfigure[t=7.5 days]{\includegraphics[height=1.7in]{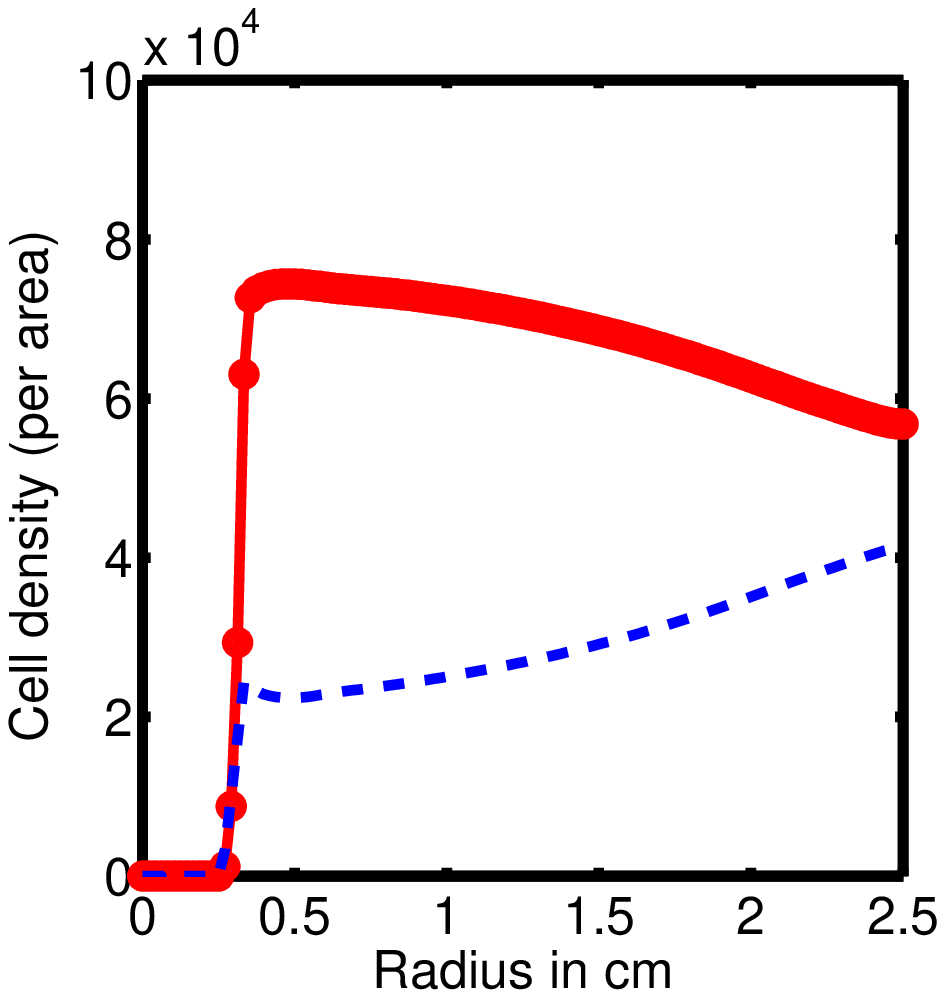}\label{dep-g}}
   \subfigure[t=10 days]{\includegraphics[height=1.7in]{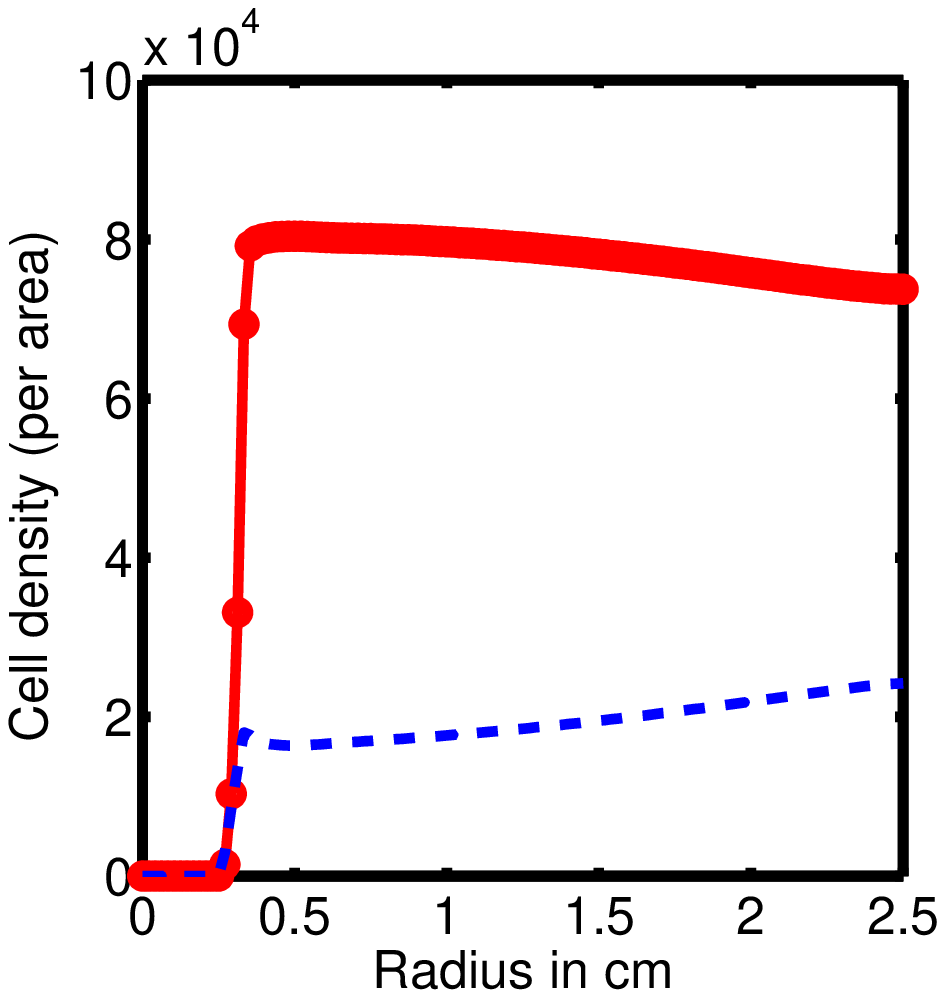}\label{dep-h}}
   \subfigure[t=20]{\includegraphics[height=1.7in]{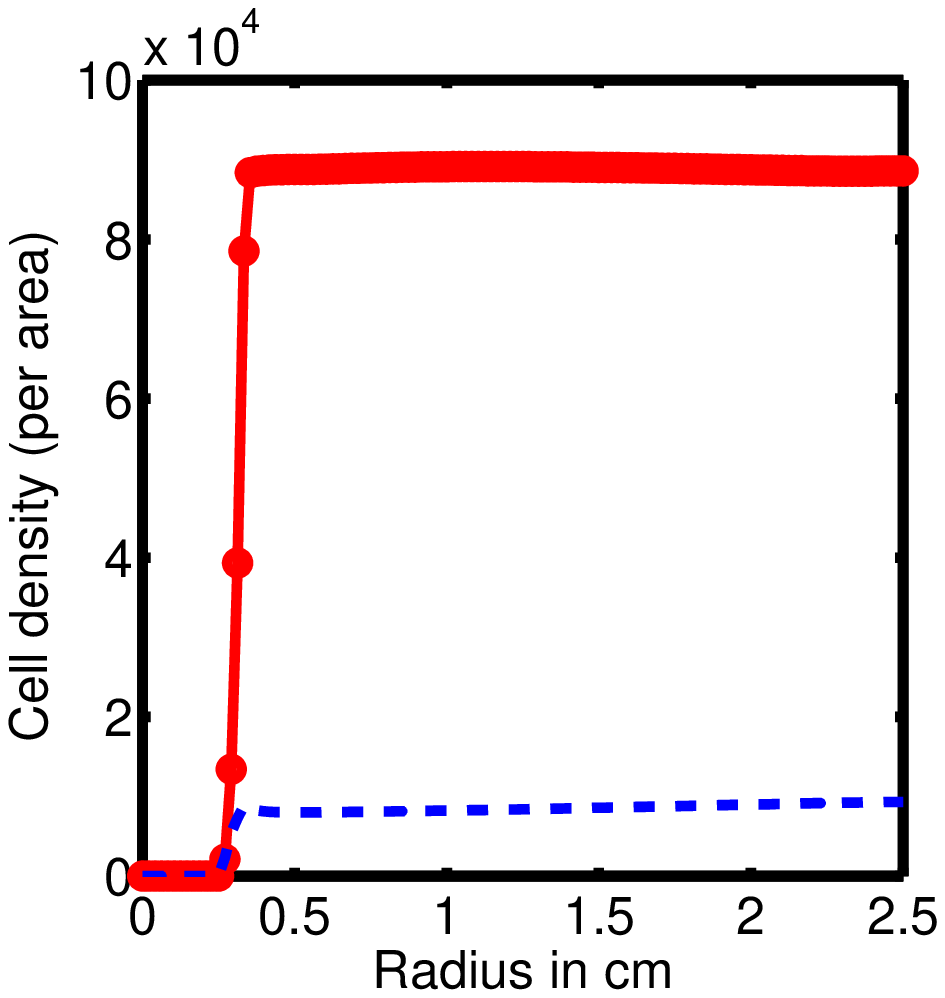}\label{dep-i}}
   \caption{Cell populations when $\sigma_{F}=0.0004$ and $\sigma_{P}=0.003$.}
   \label{wtf2}
 \end{figure}

Figure \ref{wtf2} shows the dynamics of the healthy and penumbral (sum of catabolic and EPOR active) cell populations for values
of $\sigma_{F}$ and $\sigma_{P}$ that, according to figure \ref{sigmaVary3}, lead to no radius of attenuation, that is, there is no point in our domain for which secondary TNF-$\alpha$ induced damage cannot spread. Again, we point out that this does not imply that no healing occurs. We see in figure \ref{dep-i} that after twenty days there is a significant healthy cell population despite that the penumbra spread throughout the entire domain. We again see in figure \ref{dep-i} the ``dip'' in the healthy population which is now more pronounced than in figure \ref{dee-d}.

\begin{figure}[hbtp]
  \centering
  \includegraphics[width=75mm,height=75mm]{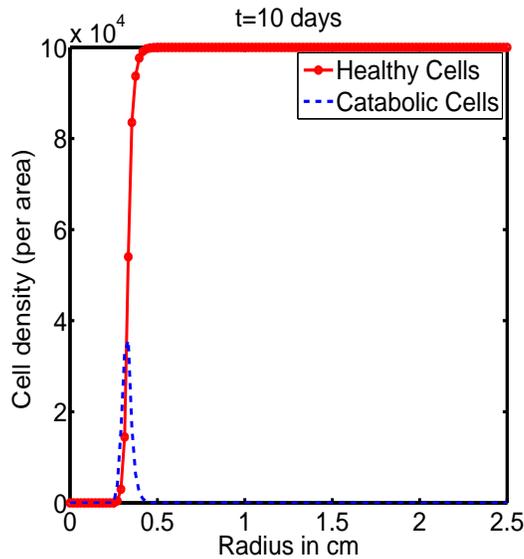}\\
  \caption{Cell populations after 10 days with $\sigma_{F}=0$. }\label{noTNF}
\end{figure}

We make one further observation. If the production rate $\sigma_{F}$ of TNF-$\alpha$ is set to zero, according to the equations in system (\ref{eq:sysa2})-(\ref{eq:sysi2}) there should be a penumbra made up entirely of catabolic cells. Figure \ref{noTNF} shows the cell population
density profile for healthy and catabolic cells after ten days. We see that it is indeed the case that there is a penumbra made up entirely of catabolic cells. This result suggests that a delay in the production of TNF-$\alpha$ could allow for the build up of a large population of
catabolic cells, so that, once TNF-$\alpha$ is produced there will be a wave of apoptosis and EPOR activation. Depending on the concentration of
EPO available this scenario could lead to more damage than is typical. Thus, damage associated with a delay in the production of TNF-$\alpha$ could potentially be worse in some cases than is seen with just the delay in the production of EPO.

\section{Conclusions}

The work in the previous section is an incomplete exploration of parameter space. We note however that it does suggest that the ratio $\frac{\sigma_{P}}{\sigma_{F}}$ between the EPO production rate and the TNF-$\alpha$ production rate plays a significant role in determining the radius of attenuation. The results shown in figures \ref{wtf1}, and \ref{wtf2} imply that EPO, or more generally the anti-inflammatory arm of cartilage injury response is robust. Even in cases where the ratio $\frac{\sigma_{P}}{\sigma_{F}}$ is small but nonzero
 inflammation does not result in the uncontrolled spread of injury as in the case when $\sigma_{P}=0$. It is ultimately desirable to
 derive theoretical results that give complete detailed knowledge of the qualitative behavior of solutions to system (\ref{eq:sysa2})-(\ref{eq:sysi2}) as a function of the parameter values. This is a difficult problem due to the number of equations and large number of parameter values. One future direction for the work presented above is its application to real experiments. It is common in orthopaedics research to perform impact experiments on large animal joints, typically bovine or porcine, or harvested human joints in attempts to replicate cell-level pathology in intra-articular fractures, see \emph{e.g.} \cite{tochigi2011,tochigi2013}. It is likely that cytokine measurements relevant to the work presented here can be made from such experimental studies. Another future direction for the work presented here and in \cite{graham1,graham2} is to include mechanical effects that are important in cartilage injury. Of particular interest is the representation of effects associated with shear stress to cartilage. In general, articular cartilage in joints such as the knees and ankles can respond efficiently to direct impact mechanical stress. However, cartilage is less resistant to shear stress. A mathematical and computational models that are capable of giving insight into what happens when shear stress is applied will be of great value to orthopaedics research.

\section{Acknowledgements}
The author would like to thank Bruce Ayati, Jim Martin, Prem Ramakrishnan, and Lei Ding for valuable discussions regarding this work. The author would like to express sincere gratitude to the editor and reviewers for their valuable comments and suggestions.

\bibliographystyle{plain}
\bibliography{epoBib}

\end{document}